\documentclass[iop]{emulateapj}

\shorttitle{Episodic High Velocity Outflows from V899 Mon: A Constrain On The Outflow Mechanisms}
\shortauthors{Ninan et al.}

\begin{document}


\title{Episodic High Velocity Outflows from V899 Mon: A Constraint On The Outflow Mechanisms\footnote{based on observations made with the Southern African Large Telescope (SALT)}}


\author{J. P. Ninan  and D. K. Ojha}
\affil{Tata Institute of Fundamental Research, Mumbai-400005, India}

\author{N. S. Philip}
\affil{St. Thomas College, Kozhencheri, Kerala, India}
\affil{Inter-University Centre for Astronomy and Astrophysics, Pune-411007, India}





\begin{abstract}
We report the detection of large variations in the outflow wind velocity from a young eruptive star, V899 Mon during its ongoing high accretion outburst phase. Such large variations in the outflow velocity (from -722 km s$^{-1}$ to -425 km s$^{-1}$) have never been reported previously in this family of objects. Our continuous monitoring of this source shows that the multi-component, clumpy, and episodic high velocity outflows are stable in the time scale of a few days, and vary over the time scale of a few weeks to months. We detect significant decoupling in the 
instantaneous outflow strength to accretion rate. From the comparison of various possible outflow mechanisms in magnetospheric accretion of young stellar objects, we conclude magnetically driven polar winds to be the most consistent mechanism for the outflows seen in V899 Mon. The large scale fluctuations in outflow over the short period makes V899 Mon the most ideal source to constrain various magnetohydrodynamics (MHD) simulations of magnetospheric accretion.   
\end{abstract}


\section{Introduction}
Young stellar objects are the drivers of the spectacular, heavily collimated outflows and jets seen in star forming regions. When matter accretes from the accretion disc to the central star, it also transfers angular momentum to it. If the angular momentum is not taken away from the system, a typical accreting classical T-Tauri star (CTTS) will reach break-up speed within the time scale of a million year. However, observationally CTTSs are found to be rotating only at $\sim$10\% of the break-up speed \citep{herbst07}. Their rotation rate is also found to be almost constant throughout their disc accretion phase spanning a few million years \citep{irwin09}. 
Outflows from these objects are the most efficient way to take away the angular momentum from the star-disc system and prevent the star from spinning up. Various theoretical outflow mechanisms which could drive outflows in young disc accreting stellar objects have been proposed in literature (see reviews by \citet{romanova15,pudritz05,konigl00}).  

EXors and FUors are a family of young low mass stars which undergo a sudden increase in the accretion rate by a factor of 10 to 100. If we assume the angular momentum removed by the outflow to be equal to the angular momentum transferred from the accretion, one can obtain a linear relation between the outflow rate and the accretion rate \citep{konigl00}. Hence, almost all theoretical models predict the outflows to be proportional to the accretion rate (they are also consistent with the observationally estimated proportionality factor $\sim$0.1).
EXor and FUor outbursts occur for a short duration (few years to decades) and many of them have associated outflows detected. Hence, they provide a unique laboratory to monitor the simultaneous evolution of the outflow with respect to the accretion rate. 
The rich optical spectrum of these sources enables us to estimate the accretion rate from the flux of a certain set of emission lines, and the outflows originating in the innermost region of the accretion disc via P-Cygni line profiles.
Previously, among other young eruptive stars, a reduction in the P-Cygni outflow absorption has been seen in V2492 Cyg when the source dimmed for a very short duration in 2010 \citep{aspin11}. However, variable extinction plays a major role in the light curve of this source \citep{hillenbrand13}, hence the correlation between the outflow and the accretion is difficult to conclude. Small scale variations in the outflow strength and velocity   
were also reported in V2493 Cyg during its rise to outburst phase by \citet{lee15}. 

Our source of this study, V899 Mon, is located near the Monoceros R2 region at a distance of $\sim$ 905 pc \citep{lombardi11}. It was detected in the quiescent phase during POSS-1 (1953) and POSS-2 (1989) surveys\footnote{Palomar Observatory Sky Survey}. The Catalina Real-time Transient Survey (CRTS) first discovered the source to be brightening into an outburst in 2005 \citep{wils09}. V899 Mon abruptly stopped its first outburst in 2010 and transitioned to a short one year quiescence in 2011. By 2012, it initiated a second outburst and since then has been in the outburst state till our last observation in 2016. V899 Mon is a flat-spectrum or an early Class II source. The mass is estimated to be around 1.5 to 3.7 M$_{\odot}$ based on the photometric and spectral energy distribution (SED) fits. In the near-infrared color-color diagram, V899 Mon falls on the classical T-Tauri locus \citep{ninan15}. A significant 20 M$_{\odot}$ clump emission is also seen in the far-infrared images, however, the optical extinction to V899 Mon was found to be quite low (A$_V$ $\sim$ 2.6 mag). In the family of young eruptive stars, \objectname{V899 Mon} showed the most dynamic changes in outflow with respect to accretion when the source transitioned from its first outburst in 2010 to a short one year quiescence in 2011 and then back to its ongoing second outburst in 2012 \citep{ninan15}. Apart from the detection of the complete disappearance of outflow P-Cygni profiles for a long duration when the accretion rate dropped to the quiescent level, we could also detect significant short timescale decoupling between the outflow wind strengths and the accretion rates. 
Non correlated variability between accretion and wind indicators is also seen in Class I protostars \citep{connelley14}. Overall, there seems to be a complex relationship between accretion and the wind in short time and length scales.

In this paper, we report the evolution of the outflow velocity from -722 km$s^{-1}$ to -425 km$s^{-1}$ in V899 Mon between our high resolution spectroscopic observations taken in 2014 December and 2015 December. Our latest observation on 2016 February shows the outflow velocity again increased to -550 km$s^{-1}$.  Such large changes in the high velocity outflow components have never been reported in this family of low mass stars. In Section \ref{sec:obs} we outline our observations and in Section \ref{sec:profiles} we report the evolution in the line profiles, and the outflow components. In Section \ref{sec:mechanisms} we compare and discuss various possible theoretical outflow mechanisms which could explain the observed episodic outflow events. We conclude in Section \ref{sec:conclusions} with the most-likely scenario of magnetic pressure driven polar stellar winds. These direct multi-epoch observations of young eruptive stars provide measurements which can be directly compared with the theoretical magnetohydrodynamic (MHD) simulations of the magnetospheric accretion and outflows.

\section{Observations} \label{sec:obs}
\subsection{Medium resolution optical spectroscopy}
Multi-epoch medium resolution (R$\sim$1000) optical spectroscopic observations of V899 Mon, during its ongoing second accretion outburst phase, were carried out using the Himalaya Faint Object Spectrograph and Camera (HFOSC) on the 2m Himalayan {\it Chandra} Telescope (HCT) at the Indian Astronomical Observatory, Hanle (Ladakh). 
Our monitoring campaign started in 2009 November 30 till 2016 February 21. Out of which, spectra up to 2014 December are already published in \citet{ninan15}.
For the remaining spectra, the spectroscopic data reduction was done using our publicly released HFOSC pipeline. The reduction procedure followed was exactly similar to \citet{ninan15}.
\subsection{High resolution optical spectroscopy}
Three epochs of optical high resolution (R$\sim$37,000) spectroscopic observations of V899 Mon, during its ongoing second outburst phase in 2015 December 15, 19, and 2016 February 20, were carried out using the High Resolution Spectrograph (SALT-HRS) \citep{bramall10} on the Southern African Large Telescope\footnote{under the program 2015-2-SCI-038 (PI: N. S. Philip)}. The instrument configuration and data reduction were carried out similar to our earlier 2014 December 22 observation using HRS reported in \citet{ninan15}. The flux calibration of the emission lines were done by scaling the normalized segment of high resolution spectrum to match the nearby epoch's flux-calibrated spectrum observed using HFOSC at HCT. From our photometry, the change in continuum flux between 2015 December 15, 19, and 2014 December 22 is found to be less than 4\% ($\leq$ 0.04 mag).

\section{Outflow Evolution} \label{sec:profiles}
\subsection{Forbidden Lines} \label{forbidlines}
\begin{figure*}
 \includegraphics[width=1\textwidth]{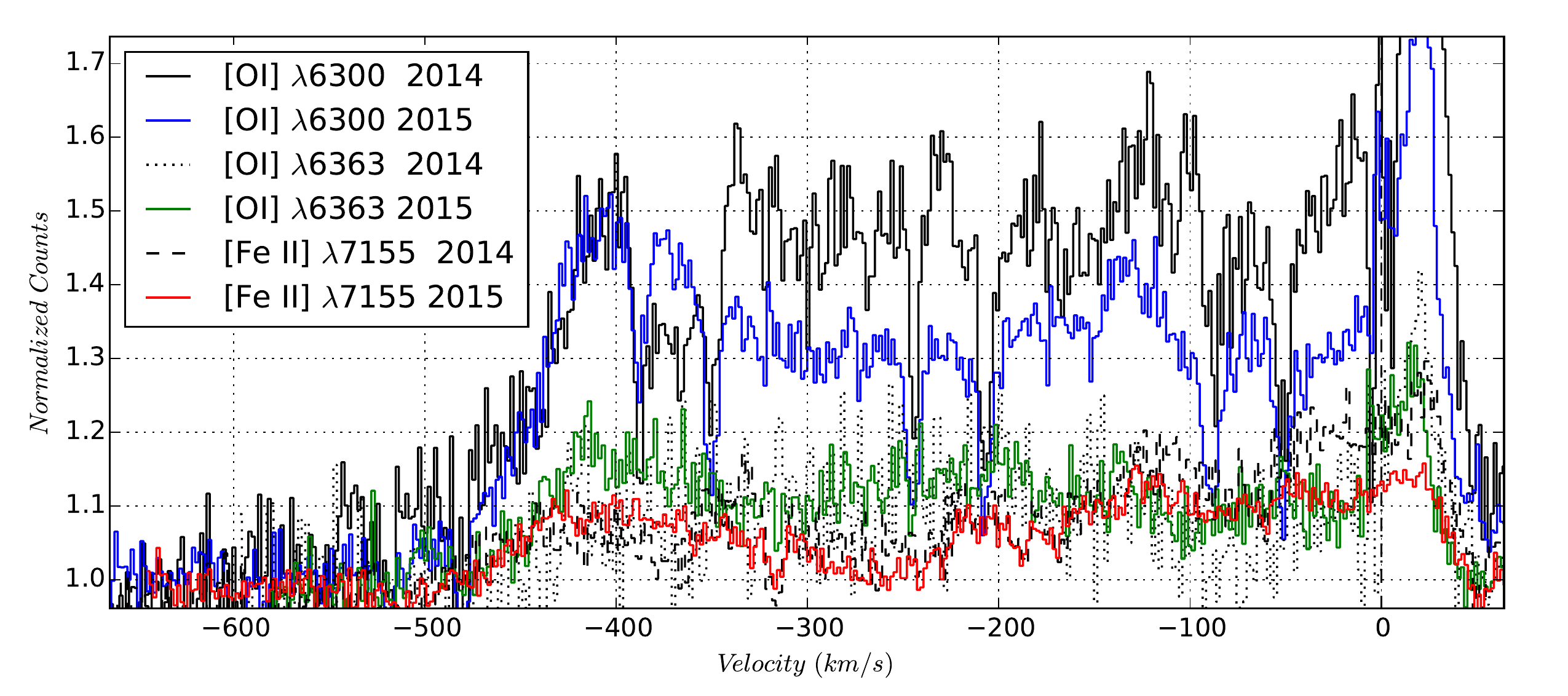}
 \caption{Evolution between 2014 December and 2015 December of the forbidden line profiles originating in the shock from V899 Mon. The high velocity components in the shock extend up to $\sim$ -475 km s$^{-1}$. The profiles do not show any significant variation between 2015 December and 2016 February, hence they are not plotted here.}
 \label{fig:forbidenProfiles}
 \end{figure*}

Figure \ref{fig:forbidenProfiles} shows the change in the profile structure of the forbidden lines [O I] $\lambda$6300, $\lambda$6363 and [Fe II] $\lambda$7155. The blue-shifted high velocity ($\sim$ -475 km s$^{-1}$) components of these three line profiles have similar velocity structure and are believed to be formed from shocks in the jet, away from the central star. Over a timescale of one year, between 2014 December and 2015 December, we could detect a drop in the medium velocity plateau component with respect to the maximum velocity component in [O I] $\lambda$6300. 
We do not see any significant short timescale variation between 2015 December 15, 19 and 2016 February 20 spectra. A hydromagnetic jet collimated by the toroidal magnetic fields does not spread sideways as fast as an ordinary hydrodynamic jet when they encounter a jet shock \citep{konigl00}. Thus the evolution of the relative column density of various projected velocity components (obtained from these blue-shifted forbidden line profiles), in principle, contain information to differentiate between the type of jets. However, such an analysis is beyond the scope of this paper. We shall use only the maximum velocity from these shock lines to infer the inclination angle and the terminal velocity in Section \ref{sec:mechanisms}.

\subsection{H$\alpha$, H$\beta$ \& Ca II IR triplet lines} \label{HaCaIIlines}
\begin{figure}
 \includegraphics[width=0.5\textwidth]{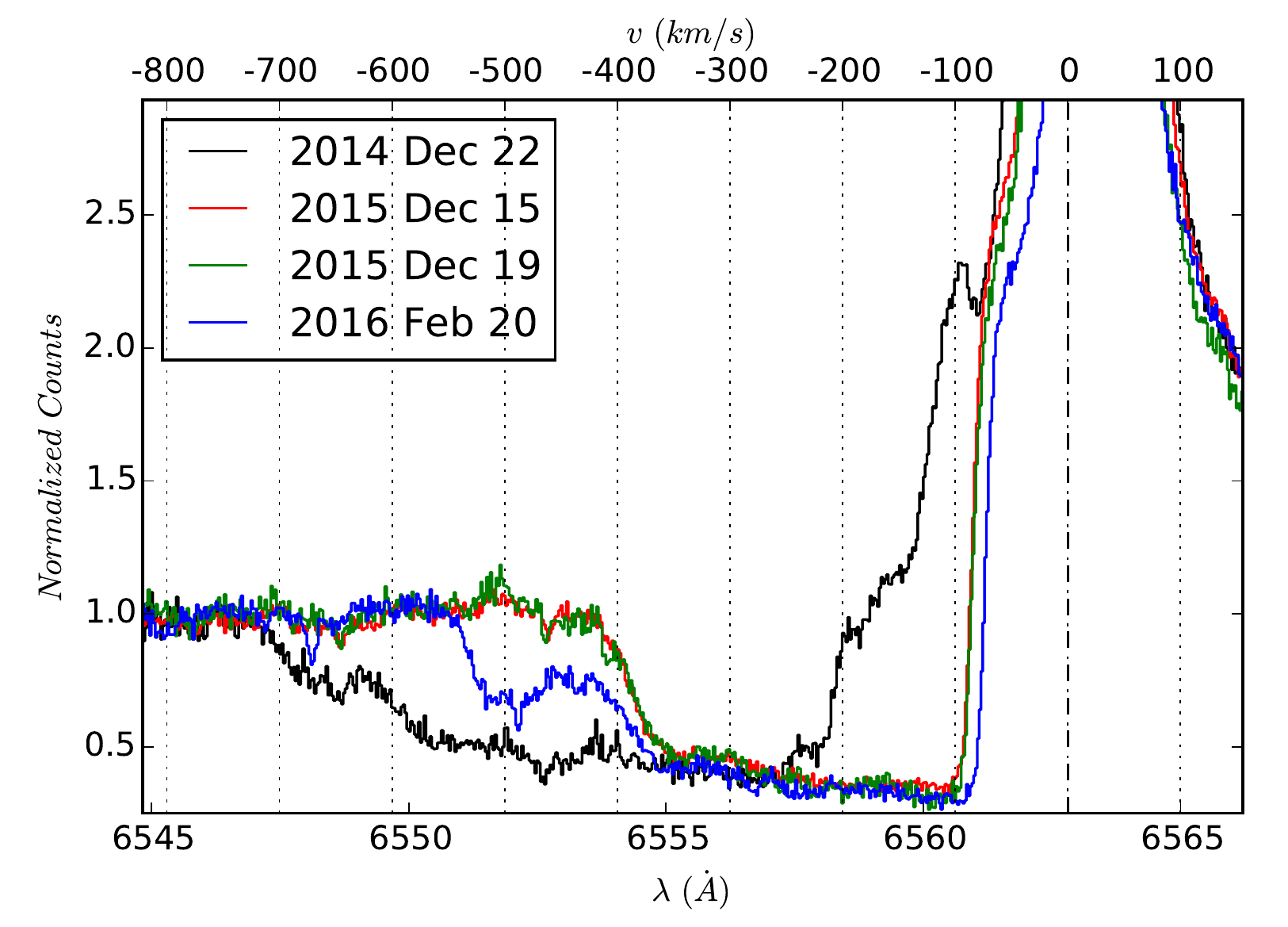}
 \caption{Outflow evolution between 2014 December 22, 2015 December 15, 2015 December 19, and 2016 February 20 of the H$\alpha$ line profile from the magnetosphere of V899 Mon. All profiles are shifted to heliocentric velocity (by 0.92 km $s^{-1}$, 4.22 km $s^{-1}$, 2.39 km $s^{-1}$ and -22.31 km $s^{-1}$ respectively).}
 \label{fig:halpha}
 \end{figure}

\begin{figure}
 \includegraphics[width=0.5\textwidth]{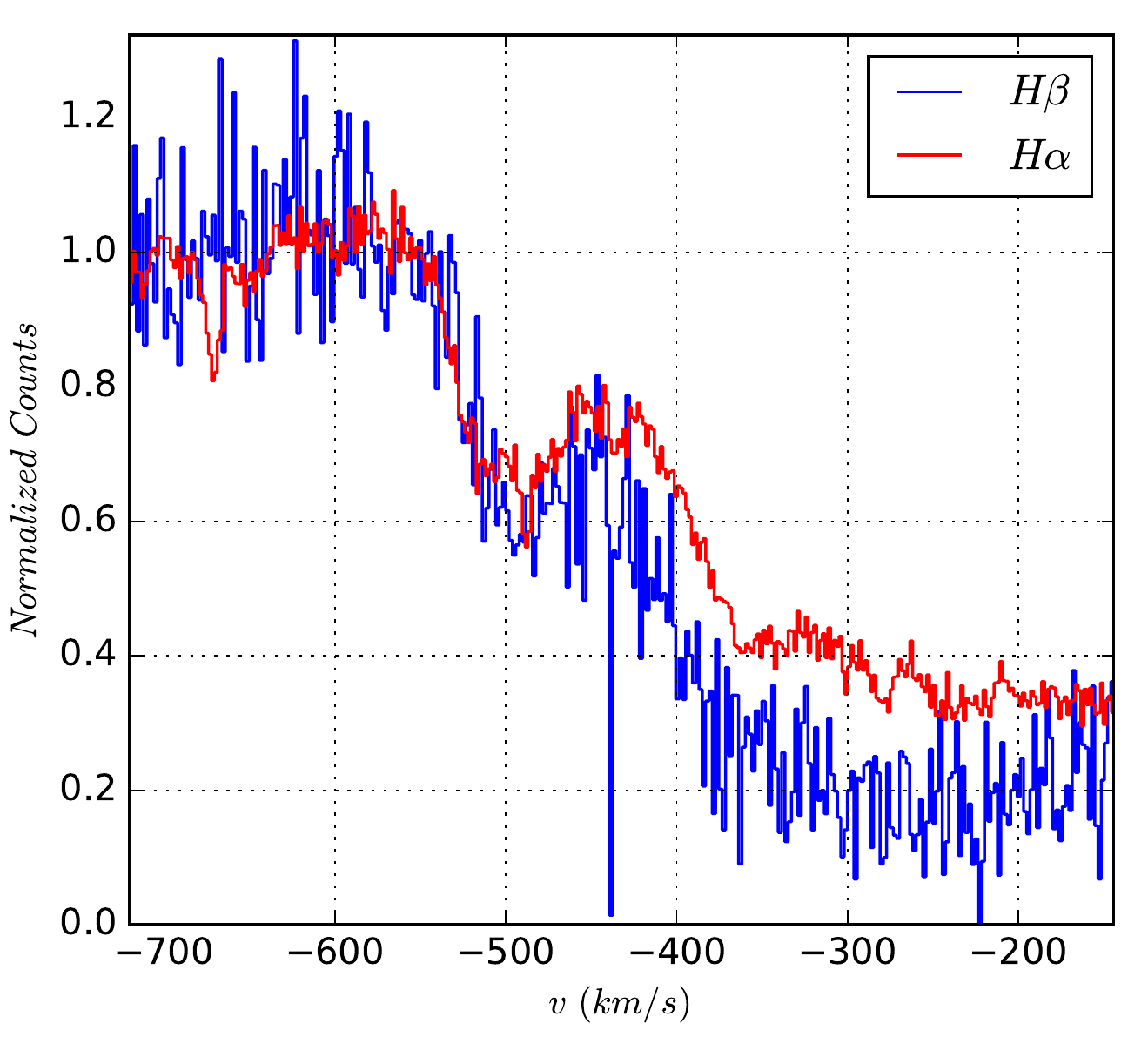}
 \caption{High velocity outflow component in 2016 February spectrum seen in both H$\alpha$ and H$\beta$ normalized line profiles of V899 Mon. Despite their different oscillator strengths, equivalent width of both H$\beta$ and H$\alpha$ of the high velocity component at -500 km s$^{-1}$ (which appeared in 2016 February) show even the high velocity outflow's H lines are saturated. }
 \label{fig:hbetaANDalpha}
 \end{figure}

Figure \ref{fig:halpha} shows the evolution in the blue-shifted absorption due to the outflow in H$\alpha$ line profile between 2014 December, 2015 December and 2016 February. The very high velocity component (extending up to -722 km s$^{-1}$) in H$\alpha$ present during 2014 December disappeared in both the 2015 December spectra. 2016 February 20 profile shows re-emergence of a high velocity component in outflow extending up to -550 km s$^{-1}$. Figure \ref{fig:hbetaANDalpha} shows this high velocity component in the H$\beta$ profile plotted over the H$\alpha$ profile. If the high velocity clump was optically thin, the ratio of the equivalent widths would have been equal to the ratio of the oscillator strength (0.6407 for H$\alpha$ and 0.1193 for H$\beta$) times line wavelength. However, the H$\beta$ profile has same equivalent width as H$\alpha$, hence the high velocity clump is optically thick. The shallower depth of this high velocity outflow absorption with respect to the lower velocity absorption implies the high velocity winds are hotter at optical depth $\tau$ $\sim$ 2/3. Hence, its density and/or temperature are different from the low velocity outflow components.

\begin{figure*}
   \centering
  \begin{tabular}[b]{@{}p{0.45\textwidth}@{}}
    \centering \includegraphics[width=0.5\textwidth]{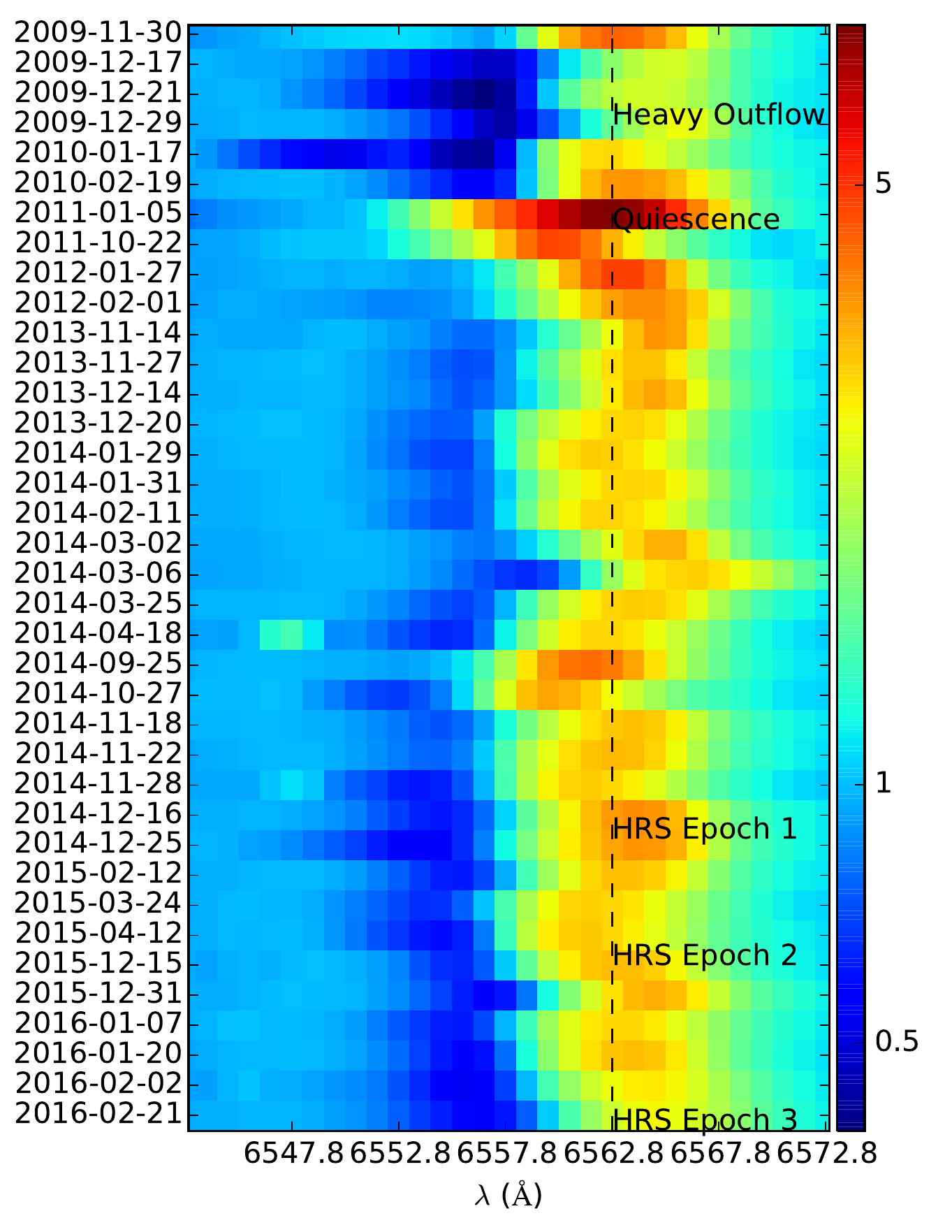} \\
    \centering\small (a)
  \end{tabular}%
  \quad
  \begin{tabular}[b]{@{}p{0.45\textwidth}@{}}
    \centering\includegraphics[width=\linewidth]{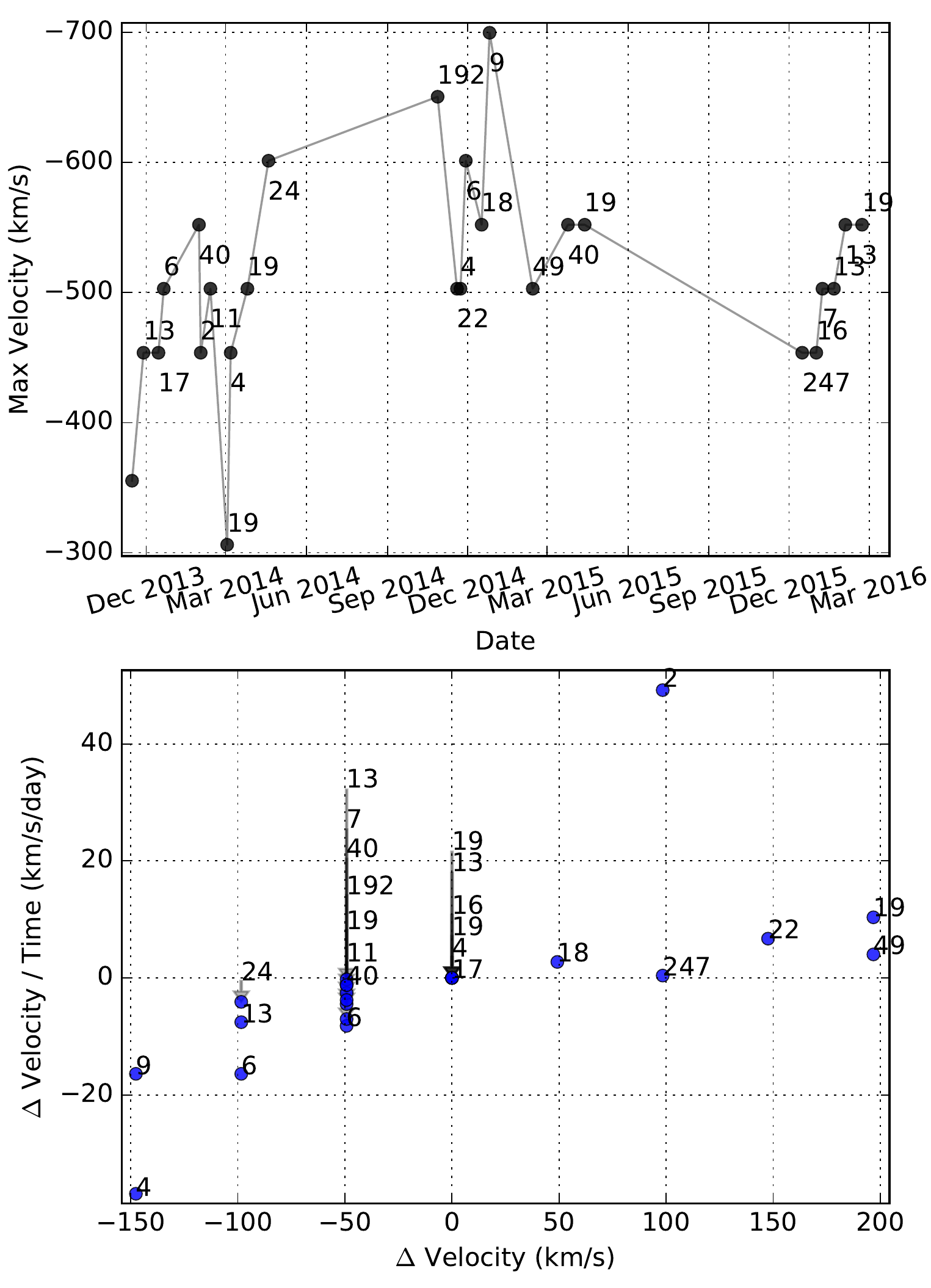} \\
    \centering\small (b)
  \end{tabular}

 \caption{Outflow evolution traced in our medium resolution spectra from 2009 to 2016 of the H$\alpha$ line profile. (a) The three epochs of HRS observations are marked in the map. The color scale is in log scale. The variation in the extent of blue-shifted absorption shows the variation in peak velocity of the outflow at each epoch. The vertical dashed line is the center wavelength of the H$\alpha$ line.  (b) The time evolution in the outflow velocity at 95\% drop from the continuum in absorption profile from the spectra taken during the ongoing second outburst. The numbers marked at each point show the number of days passed to the previous observational data. The plot at the bottom shows the scatter plot between the floor rate of change in the outflow velocity (obtained by dividing velocity change with the number of days between the consecutive observations) with respect to the change in the outflow velocity. This distribution represents both the magnitude and the timescales of the fluctuations in velocity.}
 \label{fig:halphaHFOSC}
 \end{figure*}

We did not detect any significant change in the outflow structure in timescale of 4 days between 2015 December 15 and 19. Even though we could not resolve the detailed structures, very high velocity episodic outflow components have also been previously detected at various epochs (especially at the end of the first outburst) in our medium resolution spectra \citep{ninan15}. We used our new medium resolution spectra of the second outburst to obtain the typical timescales of the variation in outflow. Figure \ref{fig:halphaHFOSC} shows the large variation in outflow velocity component detected in all of our medium resolution spectra. The velocity plotted in Figure \ref{fig:halphaHFOSC}(b) (upper panel) corresponds to 95\% drop in the blue-shifted absorption profile from the continuum level during the second outburst. It shows significant velocity variations in timescales as short as a few days. Hence, the large variation we detected in the high resolution spectra was the norm during the ongoing second outburst. As far as we know, such high velocity changes have never been reported in this family of heavily disc accreting young objects. The bottom panel of Figure \ref{fig:halphaHFOSC}(b) shows a scatter plot of  the rate of change in the outflow velocity versus the change in the outflow velocity. This scatter plot distribution, which contains information regarding both the magnitude as well as the rate of change in the outflow velocity, could be used to quantitatively match a similar scatter plot distribution of the outflow velocity fluctuations in MHD simulations.

\begin{figure*}
  \centering
  \begin{tabular}[b]{@{}p{0.45\textwidth}@{}}
    \centering\includegraphics[width=\linewidth]{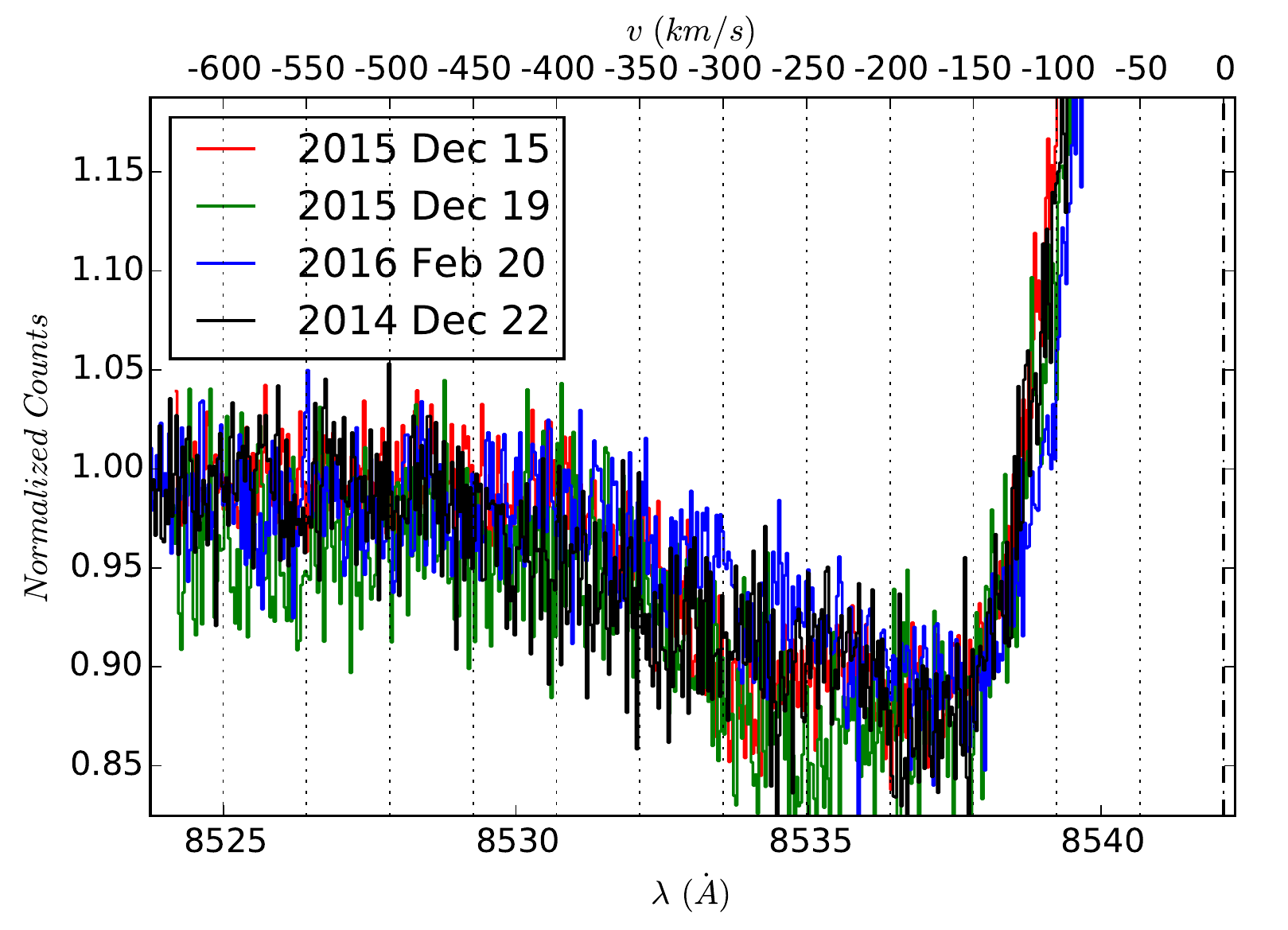} \\
    \centering\small (a) $\lambda$8542
  \end{tabular}%
  \quad
  \begin{tabular}[b]{@{}p{0.45\textwidth}@{}}
    \centering\includegraphics[width=\linewidth]{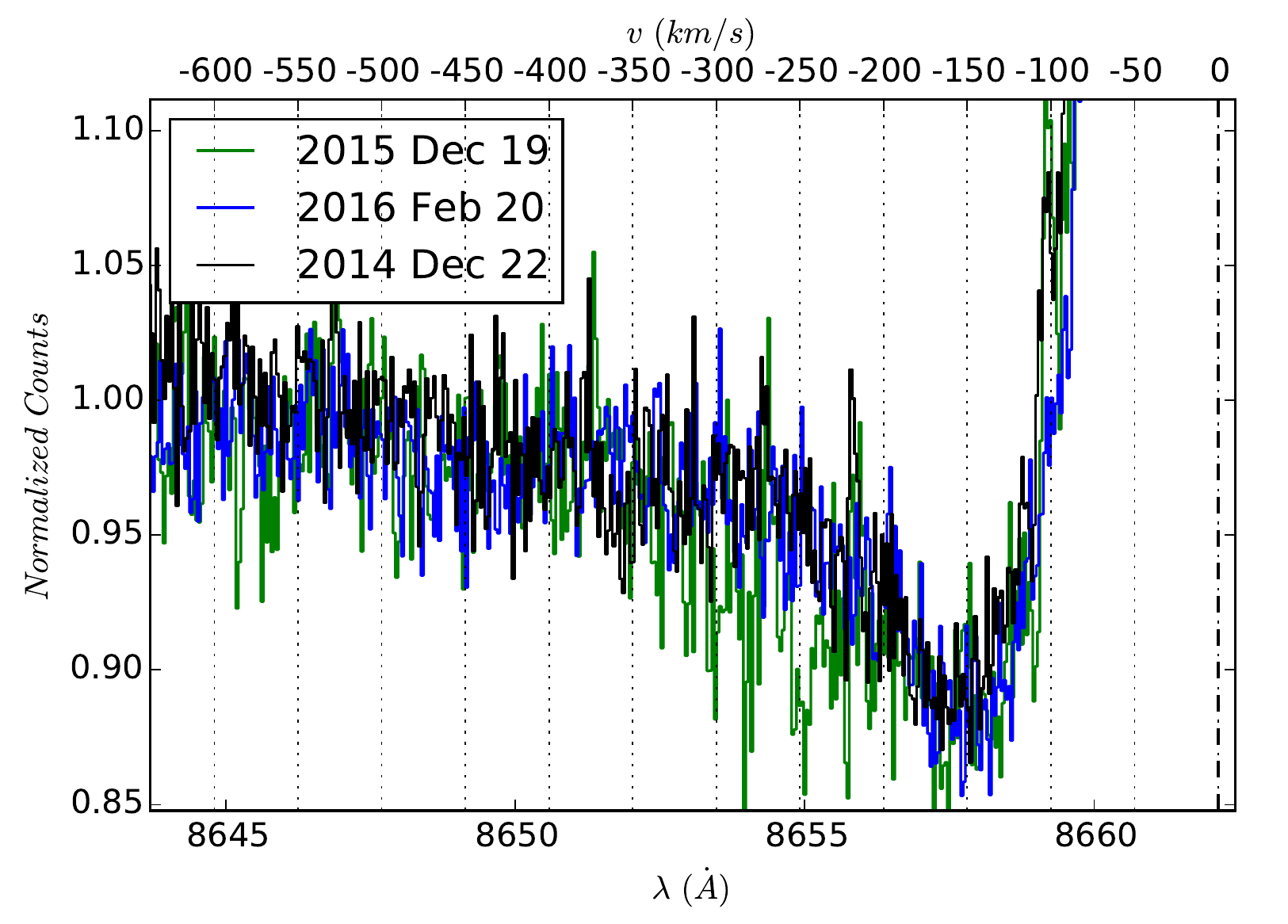} \\
    \centering\small (b) $\lambda$8662
  \end{tabular}
  \caption{Outflow evolution between 2014 December, 2015 December and 2016 February traced by Ca II IR triplet P-Cygni absorption line profiles of V899 Mon. We do not see any significant variation in these optically thin absorption profiles.}
  \label{fig:caII}
\end{figure*}

\begin{figure}
 \includegraphics[width=0.5\textwidth]{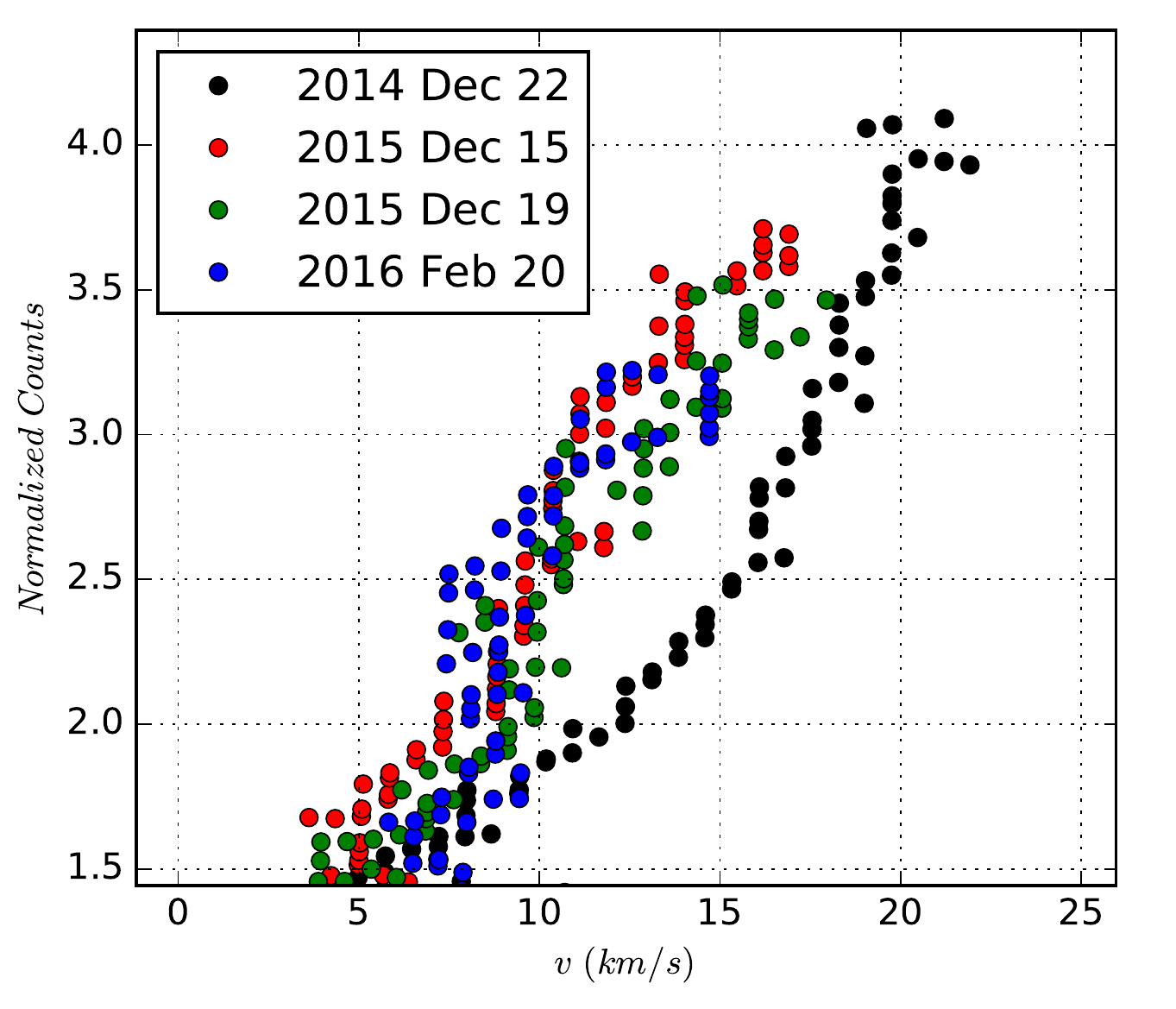}
 \caption{Evolution in the line bisector of the optically thick Ca II IR triplet $\lambda$8498 emission line. This line is least affected by the absorption from outflow. Each of the color points are the bisectors of the corresponding night's line profile. Since the line bisector is taken on an optically thick emission line, the y axis corresponds to the normalized flux counts of the line profile. The extreme flat wings of the profile in the range 1 to 1.5 is discarded from the bisector analysis due to the large uncertainties in the bisector values.}
 \label{fig:CaIILineBisector}
 \end{figure}

Outflow signatures are also seen in the  high resolution P-Cygni profiles of Ca II IR triplet lines (Figure \ref{fig:caII}). These absorption components are optically thin; interestingly they did not show any high velocity wind component which was seen in the H$\alpha$ profile during 2014 December or 2016 February.  The Ca II IR triplet line $\lambda$8498 is least affected by the blue-shifted absorption from outflow. Hence, we did the line bisector analysis on the emission line profile of  $\lambda$8498. Figure \ref{fig:CaIILineBisector} shows the line bisectors obtained at various epochs. Since the Ca II IR triplet emission lines of V899 Mon are optically thick, the line center becomes optically thick at an outer radius region than the line wings, which become optically thick in a region more deeper and closer to the central source. The fitted line bisector shows a clear increasing blue-shift in the velocity of the gas from the regions closer to the central source. There is a significant difference between 2014 December and 2015 December spectra. This could be due to a combined effect of change in optical depth and/or velocity change.

\section{Possible Outflow Mechanisms}\label{sec:mechanisms}

\begin{figure}
 \includegraphics[width=0.5\textwidth]{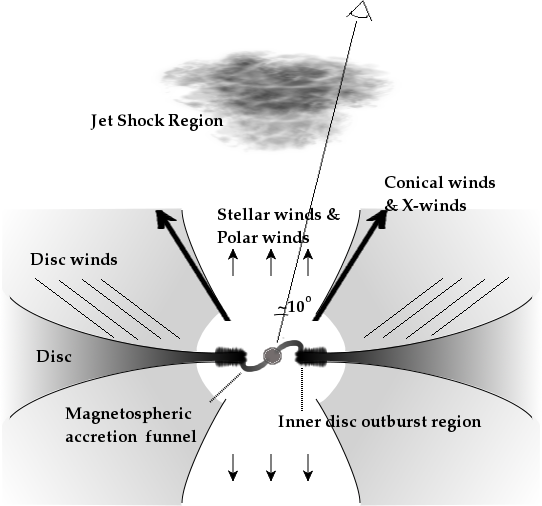}
 \caption{Schematic diagram of various outflow mechanisms which could be present in V899 Mon (see the text).}
 \label{OutflowsDiagram}
 \end{figure}

Figure \ref{OutflowsDiagram} shows the various outflow mechanisms which could operate in a young stellar object undergoing accretion from the disc via magnetospheric accretion. Disc winds are driven by magneto-centrifugal forces along poloidal magnetic field lines if they are tilted by more than 30$^{o}$ by the Blandford \& Payne mechanism \citep{blandford82}. Centrifugally driven X-winds can arise from the inner disc region if the inner disc truncation happens at the co-rotation radius of the magnetosphere \citep{shu94}. Magnetic pressure driven conical winds can also arise from the inner disc-magnetosphere boundary irrespective of the truncation radius \citep{romanova09,lii12}. Along the polar direction, one can have radiatively driven stellar winds, as well as magnetic pressure driven winds, like in the case of Poynting jet in propeller regime \citep{lovelace02,ustyugova06} when the magnetosphere is rotating fast enough.
All the above mentioned outflow mechanisms have their own characteristic velocity, outflow angle, as well as stability timescale. In this section we shall analyze each mechanism in the light of the episodic outflow velocity changes seen in V899 Mon.

 Since the estimated mass of V899 Mon is in the range of 1.5 - 3.7 M$_{\odot}$ \citep{ninan15}, it lies outside the upper boundary of low mass stars. Whether there is enough magnetic field strength to have a magnetospheric accretion for these mass ranges is a debated topic in literature. Even though V899 Mon seems to have a higher mass than typical low mass T-Tauri stars, optical and near-infrared spectra as well as radio continuum flux indicate it to be a low mass star than typical high mass stars like Herbig Ae \citep{ninan15}. This factor, along with its relatively low mass accretion rate compared to FUor (the rate is about the same order as of V1647 Ori, which is known to have a magnetospheric accretion) makes magnetospheric accretion a reasonable scenario for V899 Mon.

The inclination angle of the source is crucial to estimate the contribution of each type of outflow on the final P-Cygni profile seen in H$\alpha$ and Ca II IR triplet lines. V899 Mon is most-likely to be at a very low inclination angle and is seen almost face-on for the following reasons. The optical extinction (A$_V \sim 2.6$ mag) to the source is very small compared to what could be expected from its mid-infrared and far-infrared fluxes, and this could be the case if we are viewing V899 Mon through the low density outflow cavity \citep{ninan15}.  Since the cavity opening angles can be quite large, the constrain on inclination angle is quite wide. The maximum velocity in the blue-shifted high velocity component of the  forbidden lines (Figure \ref{fig:forbidenProfiles}) originating in the jet shocks has been shown to correlate with the inclination angle by \citet{appenzeller13}. The large velocity $\sim$ -475 km s$^{-1}$ seen in V899 Mon would thus imply a very low inclination angle according to Figure 3 in \citet{appenzeller13}.  It should be noted that the V899 Mon's velocity falls outside the region of their correlation plot by $\sim$ 100 km s$^{-1}$. However, the velocity seen in V899 Mon is still within the velocities expected from T-Tauri stars. Hence, it is likely to obey the same correlation, and we could constrain the inclination angle to be low for V899 Mon.  The near symmetric profiles of H$\alpha$ and Ca II IR triplet lines  during the quiescent period in 2011 of V899 Mon are also consistent with the low inclination angle line profiles modeled by \citet{kurosawa12}. Finally, the H$\alpha$ P-Cygni profile shape in the ongoing outburst phase in 2015 December 15 and 19 is very similar to a pre-FUor, V1331 Cyg, which is known to be at a very low inclination angle. Hence, the most consistent picture is obtained when the inclination angle to V899 Mon is taken to be small ($\sim$10$^o$).

Disc winds, X-winds and Conical winds can result in blue-shifted absorption only when they occult our line of sight to the central magnetosphere. They can be in the line of sight when the inclination angle is larger than the opening angle of these winds ($\gtrsim$30$^{o}$) \citep{kurosawa12}. If the inclination angle of V899 Mon is small, these winds cannot produce the absorption component in P-Cygni profiles. However, in principle, one can still obtain a P-Cygni profile from these high inclination angle winds, if the disc's continuum radiation is much larger than that of the central star. The typical ratio between the disc continuum and central star flux required for this is about a factor of 100 (as shown by \citet{tambovtseva14}). The change in the net flux from V899 Mon system between outburst and quiescence is only about a factor of 15 ($\Delta$ R mag = 2.9), hence this is very unlikely to be the case of V899 Mon. Moreover, considering the projection angle of these winds, the actual de-projected velocity of the winds will have to be unrealistically high.

Even if we consider the scenario where the inclination angle to V899 Mon is large, the centrifugally driven winds like X-wind and disc winds will have a characteristic velocity proportional to the Keplerian orbital velocity at the base of the wind. Since the Keplerian velocity is proportional to $R^{-0.5}$, the detected change in the outflow velocity from -722 km s$^{-1}$ to -425 km s$^{-1}$ would imply a change in radius of the base of the outflow by a factor of 3. These outflow mechanisms are also relatively stable over long timescales. Other mechanisms like conical winds also typically cannot result in such high velocity winds detected in V899 Mon \citep{kurosawa12}. Magnetospheric ejection winds driven by the inflation and re-connection of magnetospheric field lines to the disc have a strong episodic nature. However, its terminal velocity is again limited by the gravitational escape speed similar to X-winds \citep{zanni13}.  Hence, if V899 Mon's inclination angle is large,  among various possible outflow mechanisms, wide angle conical winds in propeller regime is the most likely outflow mechanism. These winds have difficulty in attaining the observed high velocities, but the timescales of the stability in these winds are consistent with our observations \citep{ustyugova06,romanova09}.

The high resolution P-Cygni profile structure in H$\alpha$ line of V899 Mon during 2015 December was strikingly similar to that of V1331 Cyg, whose inclination angle is known to be small \citep{petrov14}. \citet{petrov14} could obtain a good fit to this profile in V1331 Cyg using a radiative transfer model including stellar wind. 
While a stellar wind inside the polar opening angle can explain the profile structure seen in 2015 December, we should also be able to explain both the variation in the outflow strength as well as outflow velocity seen over the timescale of weeks in V899 Mon. Even though the outflow strength dropped below the detection limit when the accretion rate dropped during the 2011 quiescence, in the ongoing second outburst, we do not see any significant causal connection between the variations in outflow strength and the accretion (estimated from emission line strengths). Smaller changes in outflow velocity ($\sim$ 50 to 100 km s$^{-1}$) could be explained as the traced velocity at the corresponding radius where the accelerating wind gets optically thick. Change in opening angle due to magnetic field strength variation in the magnetosphere will result in change in density of the outflow and thereby the radius at which wind gets optically thick \citep{petrov14}. However, this change in apparent outflow velocity has an obvious upper limit of terminal velocity. So we need to look for some other mechanisms to explain the very high velocity ($>$ -722 km s$^{-1}$) episodic outflows detected in 2014 December and other epochs during the ongoing outburst. The change in outflow strength can also be caused by any change in the fraction of accreting matter which gets redirected in the polar direction as an accretion powered stellar wind. Recent simulations by \citet{ustyugova06,romanova09,kurosawa12} have shown that it is possible to have magnetic pressure driven polar winds which redirect matter from the upper part of the magnetospheric accretion funnel flow. They are similar to the propeller regime high velocity outflow in the polar direction seen in a fast rotating magnetosphere which has a disc truncation radius outside the co-rotation radius. Such a regime is possible in young Class I stellar sources which are typically found to rotate with a period of a few days. The acceleration to high velocities like 500-1000 km  s$^{-1}$ occurs only very near to the magnetosphere where the magnetic pressure dominates and these winds later slow down as they move farther away.  Their velocity is sensitive to the accretion rate and the outflow density, and their episodic nature with a timescale of a few weeks seen in simulations \citep{romanova09,ustyugova06} are also consistent with our observations. It should be noted that \citet{kurosawa12} did not obtain any significant contribution from these polar winds to the blue-shifted P-Cygni absorption. The polar wind outflow rate was only about 10$^{-10}$ M$_{\odot}$ yr$^{-1}$ in their MHD simulations. Stellar winds were also not included in their model. This low density of polar outflow did not result in any significant absorption, however, \citet{romanova09} showed the polar high velocity jets can have up to 10\% of the conical wind mass flux. The total mass outflow from V899 Mon is $\sim$10$^{-7}$ M$_{\odot}$ yr$^{-1}$ \citep{ninan15}, hence it is possible to have mass outflow up to $\sim$10$^{-8}$ M$_{\odot}$ yr$^{-1}$ in the polar axial winds; which is sufficient to produce blue-shifted absorption in H$\alpha$ profile. Presence of stellar wind could also further increase the mass flux along polar winds \citep{romanova09}. The outflow mechanism in these simulations of the high velocity clumpy episodic outflows along the polar direction is the most observationally consistent mechanism which can explain the fluctuating high velocity winds seen in V899 Mon.

\section{Conclusions}\label{sec:conclusions}
Our multi-epoch high resolution spectroscopic observations of the outflow traced by the P-Cygni profile in V899 Mon show change in velocity from -722 km s$^{-1}$ in 2014 December to -425 km s$^{-1}$ in 2015 December and to -550 km s$^{-1}$ in 2016 February. We do not detect any significant variation in the outflow profile over a time scale of 4 days between 2015 December 15 and 19. However, our more frequently sampled medium resolution spectra show high velocity episodic outflow variations in the timescale as short as a week at other epochs of the ongoing outburst.  In the likely scenario where V899 Mon is seen at a low inclination angle to our line of sight, the most consistent mechanism which can give rise to such unstable, clumpy, short duration high velocity winds are magnetically accelerated polar winds seen in simulations by \citet{kurosawa12,romanova09}. The highly variable nature of outflow strength and velocities make V899 Mon the most dynamic source which can be used to model and constrain the time evolution in various magnetohydrodynamic simulations of the magnetospheric accretion and outflow in young stellar objects.

\acknowledgments
The authors are thankful to the anonymous referee for insightful comments and suggestions that helped to improve the manuscript.
The authors thank the staff of CREST at Bangalore and HCT at Hanle (Ladakh), operated by the Indian Institute of Astrophysics, Bangalore, for their assistance and support during the observations with HCT. The high resolution spectrum reported in this paper was obtained with the Southern African Large Telescope (SALT), and we would like to thank Dr. Brent Miszalski and the entire SALT team for conducting SALT observations. We also thank Inter-University Centre for Astronomy and Astrophysics, India, for time on SALT.

{\it Facilities:} \facility{HCT (HFOSC)}, \facility{SALT (HRS)}.

\bibliography{JPNinanBibtext}

\begin{thebibliography}{}
\expandafter\ifx\csname natexlab\endcsname\relax\def\natexlab#1{#1}\fi

\bibitem[{Appenzeller \& Bertout(2013)}]{appenzeller13}
Appenzeller, I., \& Bertout, C. 2013, Astronomy \& Astrophysics, 558, A83

\bibitem[{Aspin(2011)}]{aspin11}
Aspin, C. 2011, The Astronomical Journal, 141, 196

\bibitem[{Blandford \& Payne(1982)}]{blandford82}
Blandford, R.~D., \& Payne, D.~G. 1982, Monthly Notices of the Royal
  Astronomical Society, 199, 883

\bibitem[{{Bramall} {et~al.}(2010){Bramall}, {Sharples}, {Tyas}, {Schmoll},
  {Clark}, {Luke}, {Looker}, {Dipper}, {Ryan}, {Buckley}, {Brink}, \&
  {Barnes}}]{bramall10}
{Bramall}, D.~G., {Sharples}, R., {Tyas}, L., {et~al.} 2010, in Society of
  Photo-Optical Instrumentation Engineers (SPIE) Conference Series, Vol. 7735,
  Society of Photo-Optical Instrumentation Engineers (SPIE) Conference Series,
  4

\bibitem[{Connelley \& Greene(2014)}]{connelley14}
Connelley, M.~S., \& Greene, T.~P. 2014, The Astronomical Journal, 147, 125

\bibitem[{{Herbst} {et~al.}(2007){Herbst}, {Eisl{\"o}ffel}, {Mundt}, \&
  {Scholz}}]{herbst07}
{Herbst}, W., {Eisl{\"o}ffel}, J., {Mundt}, R., \& {Scholz}, A. 2007,
  Protostars and Planets V, 297

\bibitem[{Hillenbrand {et~al.}(2013)Hillenbrand, Miller, Covey, Carpenter,
  Cenko, Silverman, Muirhead, Fischer, Crepp, Bloom, \&
  Filippenko}]{hillenbrand13}
Hillenbrand, L.~A., Miller, A.~A., Covey, K.~R., {et~al.} 2013, The
  Astronomical Journal, 145, 59

\bibitem[{{Irwin} \& {Bouvier}(2009)}]{irwin09}
{Irwin}, J., \& {Bouvier}, J. 2009, in IAU Symposium, Vol. 258, The Ages of
  Stars, ed. E.~E. {Mamajek}, D.~R. {Soderblom}, \& R.~F.~G. {Wyse}, 363--374

\bibitem[{Konigl \& Pudritz(2000)}]{konigl00}
Konigl, A., \& Pudritz, R.~E. 2000, Protostars and Planets IV, 759

\bibitem[{Kurosawa \& Romanova(2012)}]{kurosawa12}
Kurosawa, R., \& Romanova, M.~M. 2012, Monthly Notices of the Royal
  Astronomical Society, 426, 2901

\bibitem[{Lee {et~al.}(2015)Lee, Park, Green, Cochran, Kang, Lee, \&
  Sung}]{lee15}
Lee, J.-E., Park, S., Green, J.~D., {et~al.} 2015, The Astrophysical Journal,
  807, 84

\bibitem[{Lii {et~al.}(2012)Lii, Romanova, \& Lovelace}]{lii12}
Lii, P., Romanova, M., \& Lovelace, R. 2012, Monthly Notices of the Royal
  Astronomical Society, 420, 2020

\bibitem[{Lombardi {et~al.}(2011)Lombardi, Alves, \& Lada}]{lombardi11}
Lombardi, M., Alves, J., \& Lada, C.~J. 2011, Astronomy \& Astrophysics, 535,
  A16

\bibitem[{Lovelace {et~al.}(2002)Lovelace, Li, Koldoba, Ustyugova, \&
  Romanova}]{lovelace02}
Lovelace, R. V.~E., Li, H., Koldoba, A.~V., Ustyugova, G.~V., \& Romanova,
  M.~M. 2002, The Astrophysical Journal, 572, 445

\bibitem[{Ninan {et~al.}(2015)Ninan, Ojha, Baug, Bhatt, Mohan, Ghosh,
  Men’shchikov, Anupama, {M. Tamura}, \& Henning}]{ninan15}
Ninan, J.~P., Ojha, D.~K., Baug, T., {et~al.} 2015, The Astrophysical Journal,
  815, 4

\bibitem[{Petrov {et~al.}(2014)Petrov, Kurosawa, Romanova, Gameiro, Fernandez,
  Babina, \& Artemenko}]{petrov14}
Petrov, P.~P., Kurosawa, R., Romanova, M.~M., {et~al.} 2014, Monthly Notices of
  the Royal Astronomical Society, 442, 3643

\bibitem[{Pudritz \& Banerjee(2005)}]{pudritz05}
Pudritz, R.~E., \& Banerjee, R. 2005, in , eprint: arXiv:astro-ph/0507268,
  163--173

\bibitem[{Romanova \& Owocki(2015)}]{romanova15}
Romanova, M.~M., \& Owocki, S.~P. 2015, Space Science Reviews, 191, 339

\bibitem[{Romanova {et~al.}(2009)Romanova, Ustyugova, Koldoba, \&
  Lovelace}]{romanova09}
Romanova, M.~M., Ustyugova, G.~V., Koldoba, A.~V., \& Lovelace, R. V.~E. 2009,
  Monthly Notices of the Royal Astronomical Society, 399, 1802

\bibitem[{Shu {et~al.}(1994)Shu, Najita, Ostriker, Wilkin, Ruden, \&
  Lizano}]{shu94}
Shu, F., Najita, J., Ostriker, E., {et~al.} 1994, The Astrophysical Journal,
  429, 781

\bibitem[{Tambovtseva {et~al.}(2014)Tambovtseva, Grinin, \&
  Weigelt}]{tambovtseva14}
Tambovtseva, L.~V., Grinin, V.~P., \& Weigelt, G. 2014, Astronomy \&
  Astrophysics, 562, A104

\bibitem[{Ustyugova {et~al.}(2006)Ustyugova, Koldoba, Romanova, \&
  Lovelace}]{ustyugova06}
Ustyugova, G.~V., Koldoba, A.~V., Romanova, M.~M., \& Lovelace, R. V.~E. 2006,
  The Astrophysical Journal, 646, 304

\bibitem[{{Wils} {et~al.}(2009){Wils}, {Greaves}, {Drake}, \&
  {Catelan}}]{wils09}
{Wils}, P., {Greaves}, J., {Drake}, A.~J., \& {Catelan}, M. 2009, Central
  Bureau Electronic Telegrams, 2033, 1

\bibitem[{Zanni \& Ferreira(2013)}]{zanni13}
Zanni, C., \& Ferreira, J. 2013, Astronomy \& Astrophysics, 550, A99

\end{thebibliography}

\end{document}